 \documentclass[jgrga,pdfsync]{agu2001}

 \usepackage{epsfig}

\authorrunninghead{SHAVIV}

\titlerunninghead{COSMIC RAYS AND CLIMATE SENSITIVITY}

\authoraddr{Nir J. Shaviv,
Racah Institute of Physics, Hebrew University of Jerusalem, Jerusalem, 91904, Israel.
(shaviv@phys.huji.ac.il) }

\begin{document}


%
%

%

%
%

%
%

\title{On climate response to changes in the cosmic ray flux and radiative budget}
%

%
%


\author{Nir J. Shaviv}
\affil{Racah Institute of Physics, Hebrew University of Jerusalem, Jerusalem, 91904, Israel}

\begin{abstract}
We examine the results linking cosmic ray flux (CRF) variations to global climate change. We then proceed to study various periods over which there are estimates for the radiative forcing, temperature change and CRF variations relative to today. These include the Phanerozoic as a whole, the Cretaceous, the Eocene, the Last Glacial Maximum, the 20$^\mathrm{th}$ century, as well as the 11-yr solar cycle.  This enables us to place quantitative limits on climate sensitivity to both changes in the CRF, $\Phi_{CR}$, and the radiative budget, $F$, under equilibrium.  Under the assumption that the CRF is indeed a climate driver, we find that the sensitivity to CRF variations is consistently fitted with $\mu \equiv -\Phi_0 (dT_{global}/ d \Phi_{CR})  \approx 6.5 \pm 2.5^\circ {\mathrm K} $ (where $\Phi_0$ is the CR energy flux today). Additionally, the sensitivity to radiative forcing changes is $\lambda \equiv  \left.dT_{global}/ dF\right|_0 = 0.35 \pm 0.09  \CWm$, at the current temperature, while its temperature derivative is negligible with $(d \lambda /dT)_0 = 0.01 \pm 0.03  \iWm$. If the observed CRF/climate link is ignored, the best sensitivity obtained is $\lambda = 0.54 \pm 0.12 \CWm$ and $(d \lambda /dT)_0 = -0.02 \pm 0.05  \iWm$. The CRF/climate link therefore implies that the increased solar luminosity and reduced CRF over the previous century should have contributed a warming of $0.37 \pm 0.13 \deg$, while the rest should be mainly attributed to anthropogenic causes. Without any effect of cosmic rays, the increase in solar luminosity would correspond to an increased temperature of $0.16\pm0.04\deg$. 
\end{abstract}

%
%

%

\def\omunit{{km sec$^{-1}$ kpc$^{-1}$}}
\def\gtrsim{ \lower .75ex \hbox{$\sim$} \llap{\raise .27ex \hbox{$>$}} }
\def\lesssim{ \lower .75ex \hbox{$\sim$} \llap{\raise .27ex \hbox{$<$}} }

\def\Wm{\hskip1pt {\mathrm W}\hskip1pt {\mathrm m}^{-2}}
\def\iWm{\hskip1pt {\mathrm m}^{2}\hskip1pt {\mathrm W}^{-1}}
\def\CWm{^\circ{\mathrm K}\hskip1pt\mathrm{W^{-1}\hskip1pt m^{2}}}
\def\deg{^\circ{\mathrm K}}

\def\dTdbl{\Delta T_{\times 2}}
\def\th{$^\mathrm{th}$}

\def\width{2.4in}
\def\widthsix{3in}

\def\figureone{
 \begin{figure}
\vskip -0.5cm
\hskip -0.5cm
\center{\epsfig{file=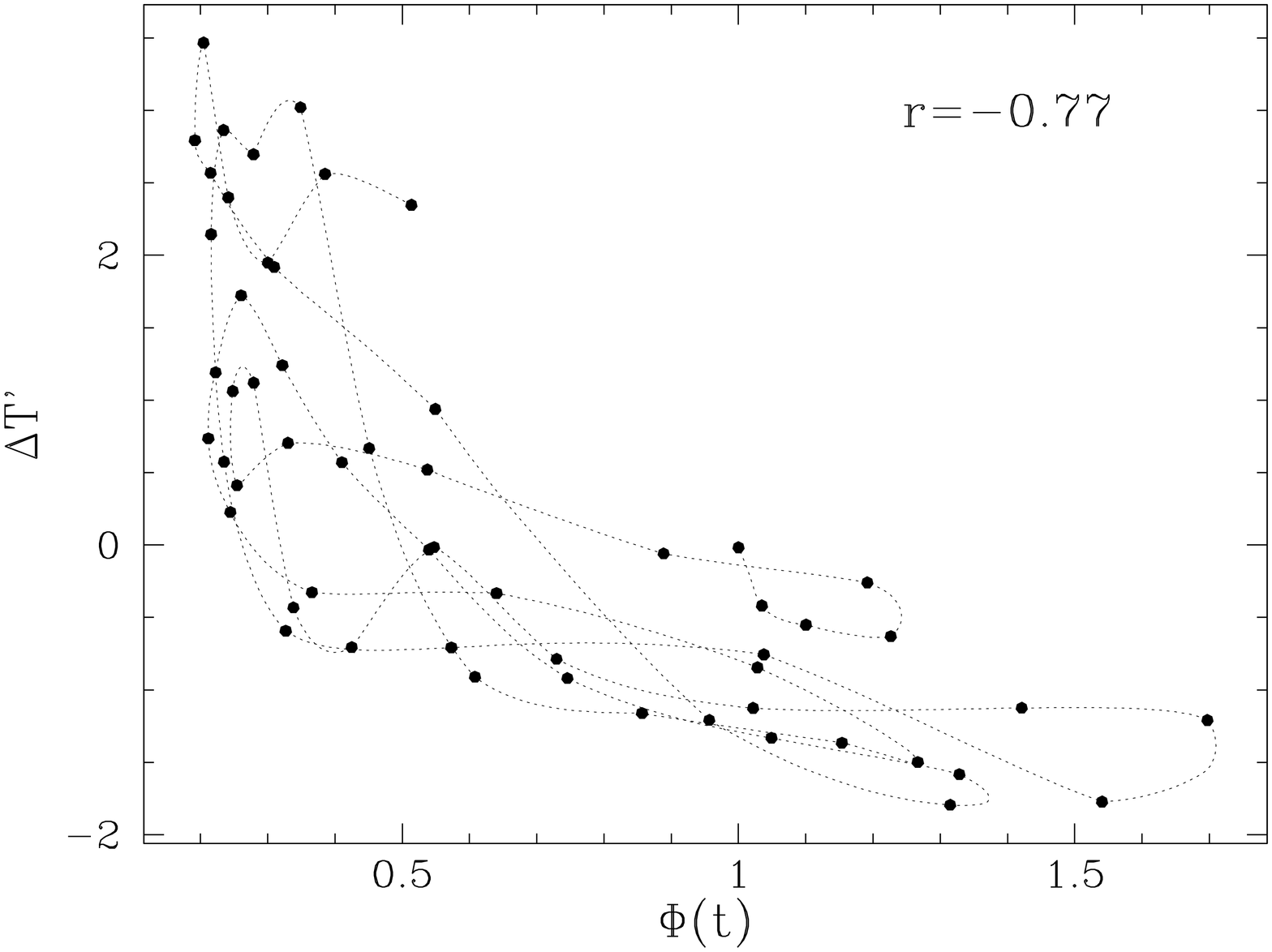,width=\width,angle=-90}}
\caption{ \small \sf 
The high correlation between the reconstructed temperature and CRF over the Phanerozoic can be used to estimate global sensitivity. Here, $\Delta T'$ is the reconstructed temperature $\Delta T$ of \cite{Veizer2000} over the past 550 Ma, with a small linear temperature increase of $1.7 \deg (t/550 Ma)$ subtracted (see \cite{Shaviv03}). The CRF is one of 3 reconstructions used in \cite{Shaviv03}. The two others differ in the total amplitude of variations. The two independent signals have a high Pearson correlation of $r=-0.77$. Although a statistically significants limits on $p$ cannot be placed, lower $p$'s are favored (with $p\sim 0.3$ producing the best fit). Nevertheless, the value of $p=0.5$ is theoretically preferred. The data is smoothed with a 50 Ma moving average. Points are 10 Ma intervals. }
 \label{fig:Phanerozoic}
 \end{figure}
 }

\def\figuretwo{
 \begin{figure}
 \vskip -1.0cm
\hskip -0.5cm
\center{\epsfig{file=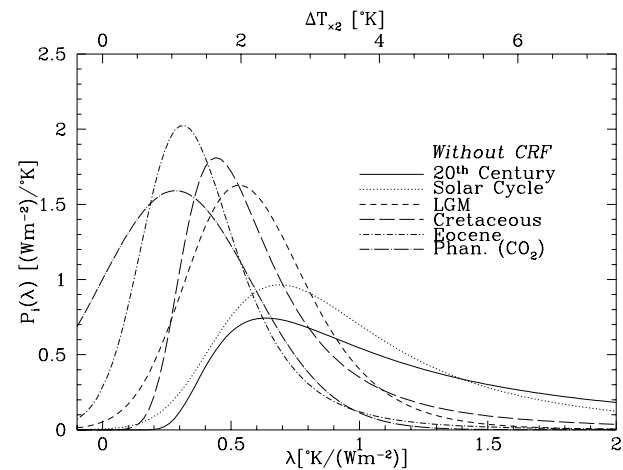,width=\width,angle=-90}}
\caption{ \small \sf 
The Probability Distribution function for $\lambda$ (and $\dTdbl$) obtained by comparing radiation budget differences to temperature change over various time scales, assuming that CRF variations {\em do not} affect the global climate (though it does includes the small solar luminosity changes). We also assume, as \cite{Gregory02}, that $\Delta T$ and $\Delta F$ entering $\lambda$ have Gaussian errors. The cases are: (1) Temperature increase over the past century (following \cite{Gregory02}).  (2) Temperature variations over 300 yrs of solar cycles.  (3) Warming since the LGM (following \cite{Hoffert92} and \cite{Hansen93}). (4) Cooling relative to the Cretaceous ($\sim 100$~Ma, \cite{Hoffert92}). (5) Cooling relative to the Eocene ($\sim 55$~Ma, following \cite{Hoffert96}). 
And (6) Phanerozoic $\Delta T$ vs.\ $\mathrm{R_{CO2}}$ (\S3.1, following \cite{Shaviv03}).
We assume that the temperature increase $\dTdbl$ following the doubling of the atmospheric CO$_2$ content corresponds to an increase of $3.71\Wm$ \citep{Myhre}.  }
 \label{fig:pdfs0}
 \end{figure}
 }

\def\figurethree{
 \begin{figure}
 \vskip -0.8cm
\hskip -0.5cm
\center{\epsfig{file=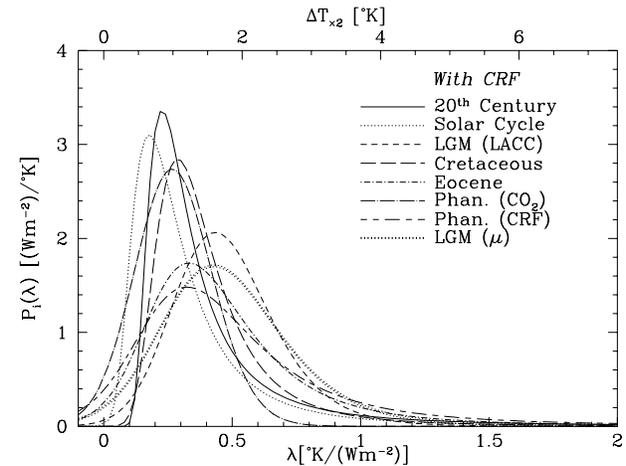,width=\width,angle=-90}}
\caption{ \small \sf 
The Probability Distribution function for $\lambda$ (and $\dTdbl$) obtained by comparing radiation budget differences to temperature change over various time scales, assuming that CRF variations {\em do} affect the global climate through modulation of the low altitude cloud cover, and that the relation between cloud cover changes and radiative forcing are given by the nominal range. The cases are the same as in figure \ref{fig:pdfs0} with the added CRF effect. In addition, we have case (7) Global sensitivity from correlation between CRF variations and temperature variations over the Phanerozoic. No equivalent for this case exists in the previous set because the CRF/climate link is required to explain this data. And (8) Warming since the LGM assuming $\mu$ is given by the CRF/temperature correlation over the Phanerozoic and not necessarily through LACC variations.
 }
  \label{fig:pdfs1}
\end{figure}
}

\def\figurefour{
\begin{figure}
\vskip -0.5cm
\hskip -0.5cm
\center{\epsfig{file=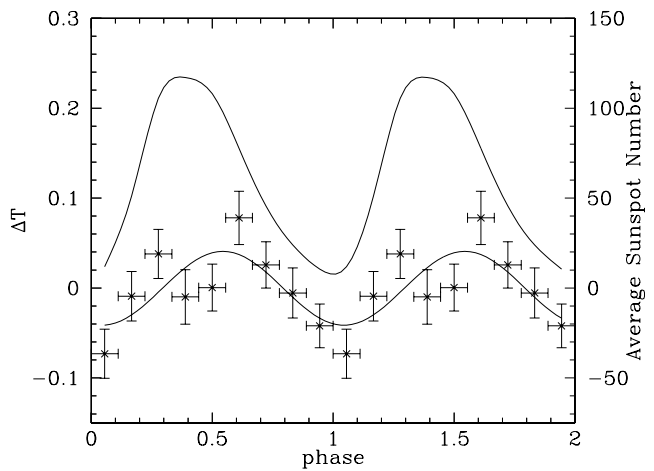,width=\width,angle=-90}}
\caption{ \small \sf 
Plotted are average number of sunspots, the global temperature and a sinosoidal fit. The global surface temperature is the reconstructed temperature over the past 300 yrs (with variations on time scales longer than 30 yrs removed) folded over the solar cycle length. The average sunspot number folded over the same cycle.
}
\label{fig:solarcycle}
\end{figure}
}

\def\figurefive{
 \begin{figure}
  \vskip -0.8cm
\hskip -0.5cm
\vskip -2cm
\center{\epsfig{file=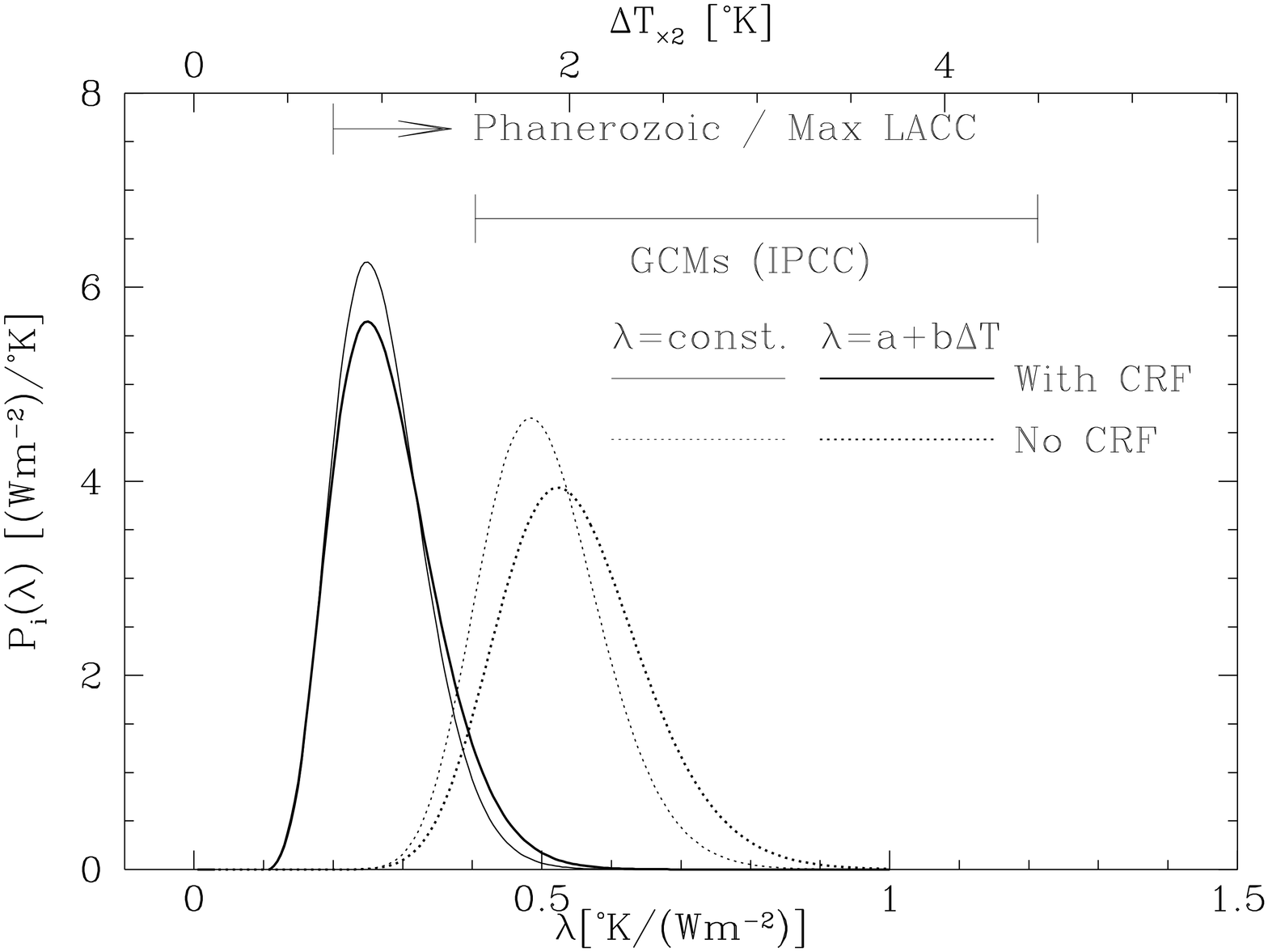,width=\width,angle=-90}}
\caption{ \small \sf 
The combined probability distribution functions for $\lambda$ obtained by combining the PDFs given in fig.~\ref{fig:pdfs0} and fig.~\ref{fig:pdfs1} when the CRF/climate link is either neglected or included. (In the latter case, the combination is done as explained in the text). Thin lines denote the result if $\lambda$ is assumed to be temperature insensitive, while the heavy line is the result obtained when $\lambda$ is allowed to be a linear function of the temperature. The the latter case, the distribution for $\lambda$ today is plotted. Also marked are the two additional constraints obtained from the Phanerozoic data which do not depend on the CRF/LACC link, as well as the sensitivity range of 1.5 to 4.5$\deg$, which according to the \cite{IPCC} is ``widely cited''. Note that the sensitivity range of the 15 GCM models actually used by the \cite{IPCC} is 2.0 to 5.1$\deg$. 
 }
 \label{fig:pdfscombined}
\end{figure}
}

\def\figuresix{
 \begin{figure*}
\hskip -0.5cm
\center{\epsfig{file=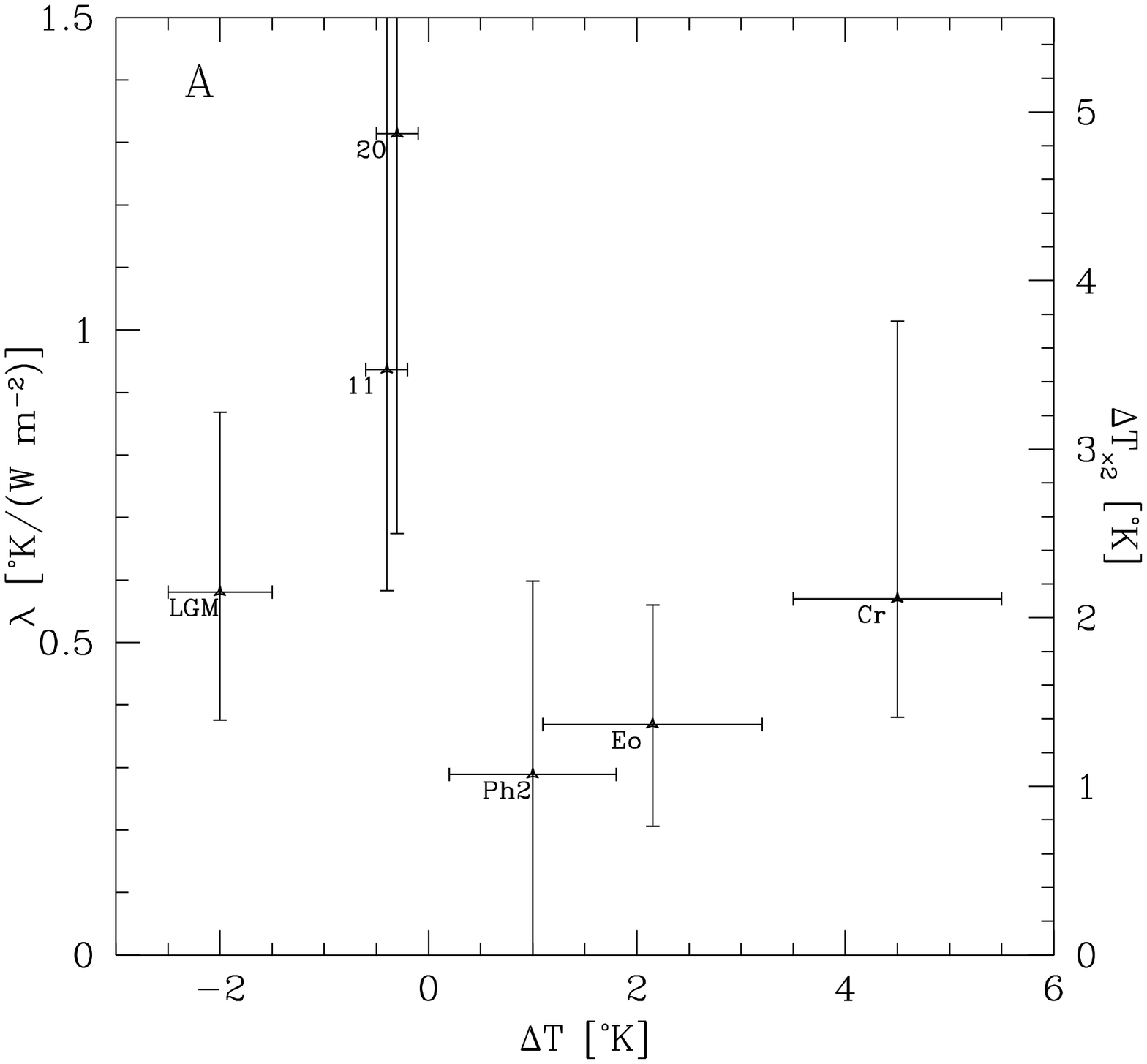,width=\widthsix} \epsfig{file=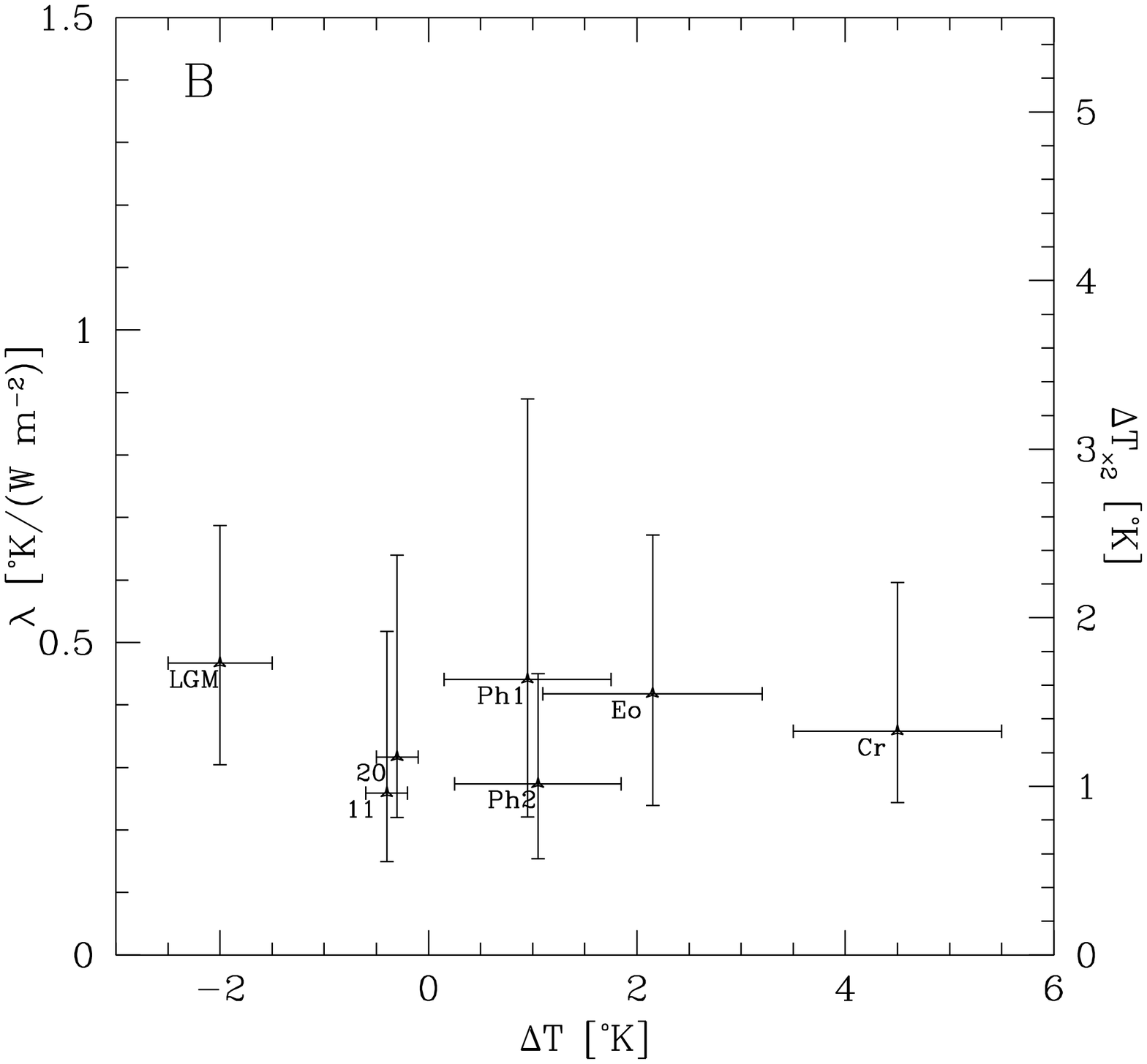,width=\widthsix}} 
\caption{ \small \sf 
The estimated sensitivity $\lambda$ as a function of the {\em average} temperature  $\Delta T$ relative to today over which the sensitivity was calculated (e.g., average temperature between today and a given epoch if conditions at the given epoch and today are used to estimate the sensitivity). The values are for the Last Glacial Maximum (LGM), 11 year solar cycle over the past 200 years (11), 20$^{\mathrm{th}}$ century global warming (20), Phanerozoic though comparison of the tropical temperature to CRF variations (Ph1) or to CO$_2$  variations (Ph2), Eocene (Eo) and Mid-Cretaceous (Cr). Panel (a)  assumes that the CRF contributes no radiative forcing while panel (b) assumes that the CRF does affect climate. Thus, the ``Ph1" measurement is not applicable and does not appear in panel (a). From the figures it is evident that: (i) The expectation value for $\lambda$ is lower if CRF affects climate. (ii) The values obtained using different paleoclimatic data are notably more consistent with each other if CRF does affect climate (iii) There is no significant trend in $\lambda$ vs.\ $\Delta T$.
 }
 \label{fig:results}
\end{figure*}
}

\def\figureseven{
 \begin{figure}
\hskip -0.5cm
\center{\epsfig{file=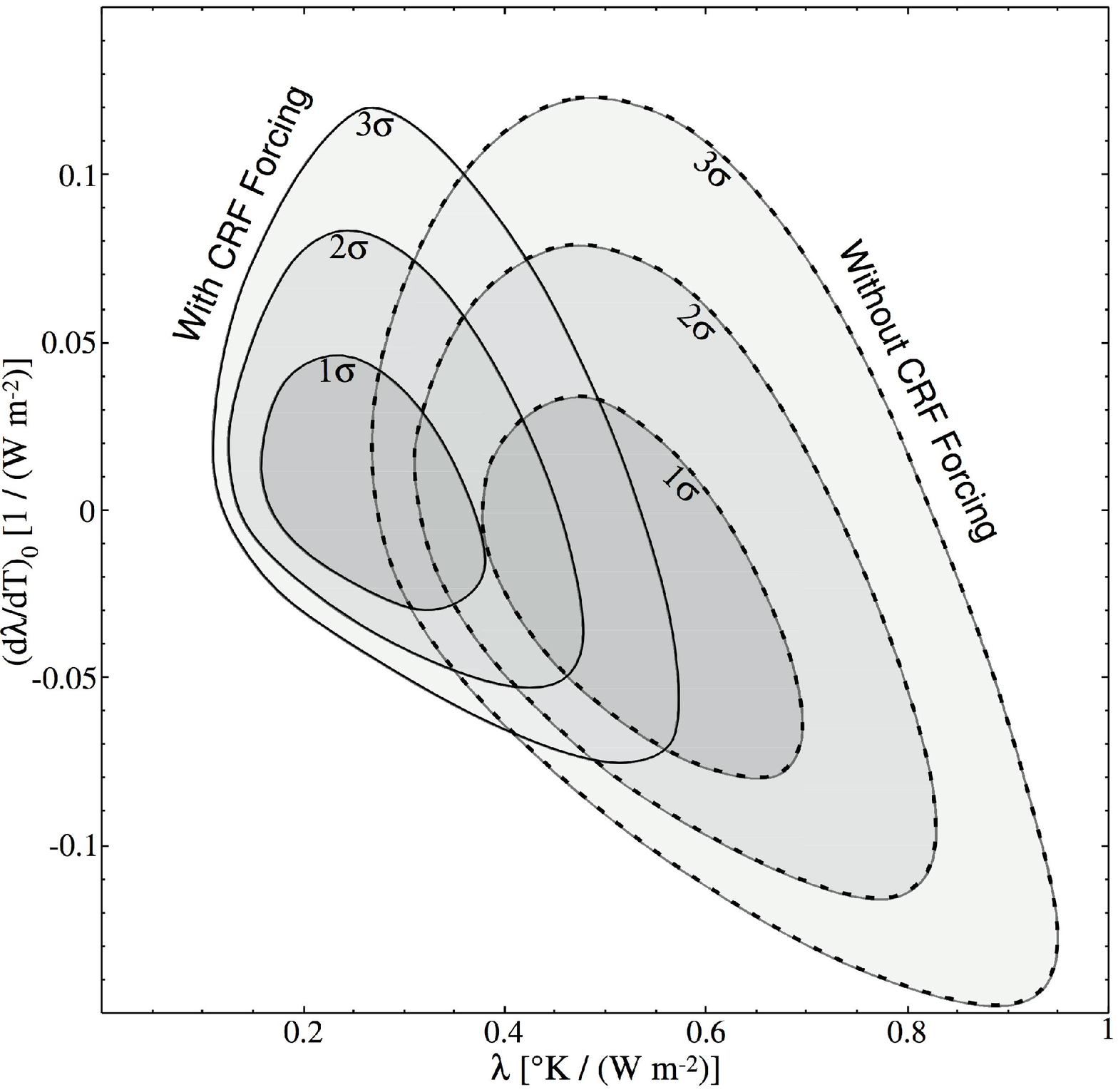,width=3in}}
\caption{ \small \sf 
The two dimensional probability distribution functions for $\lambda$ and its temperature sensitivity $(d \lambda / d t)$, assuming the sensitivity is at most linear in the temperature. It is evident that  (i) the expectation for $\lambda$ is lower once the  CRF/climate link is introduced. (ii) In both cases, there is no clear dependence of $\lambda$ on the temperature. Namely, $ | d \lambda / dT | \lesssim 0.05~(W/m^2)^{-1}$.}
 \label{fig:contours}
\end{figure}
}

\begin{article}

\section{Introduction}

Accumulating evidence suggests that solar activity is responsible for at least some climatic variability. These include correlations between solar activity and either direct climatic variables or indirect climate proxies over time scales ranging from days to millennia  \citep{Herschel,Eddy,Labitzke1992,FS,Soon1996,Soon2000,Beer2000,Maya,Oman}. It is therefore difficult at this point to argue against the existence of any causal link  between solar activity and climate on Earth.  However, the climatic
 variability attributable  to solar activity is larger than
 could be expected from the  typical 0.1\% changes in the solar
 irradiance observed over the decadal  to centennial time scale \citep{Beer2000,Soon2000}.  Thus, an amplifier is required unless the sensitivity to changes in the radiative forcing is uncomfortably high.
 
The first suggestion for an amplifier of solar activity was suggested by \cite{Ney}, who pointed out that if climate is sensitive to the amount of tropospheric ionization, it would also be sensitive to solar activity since the
solar wind modulates the cosmic ray flux (CRF), and with it, the amount of
tropospheric ionization \citep{Ney}. 

 Over the solar cycle, the solar wind
strength varies considerably, such that the amount of
tropospheric ionization changes by typically 5\%-10\%.   
\cite{Sven1998,Sven2000}, \cite{Marsh2000} as well as
\cite{SecondAnalysis} have shown
that the variations in the amount of low altitude cloud cover (LACC) 
nicely correlate with the CRF reaching Earth over two decades. Both
signals, of cloud cover and CRF variations lag by
typically half a year behind other solar activity indices. This
suggests that it is more likely that the cloud cover is directly
related to the CRF than directly to solar activity. 


More recent data on the LACC seems to exhibit a weaker correlation  with the variable CRF (e.g., \cite{Farrar}). There are however a few peculiarities in the data which are indicative of a calibration problem, which once removed, seem to recover the high correlation between the CRF and the LACC \citep{SvenCalibration}. For an objective review, the reader is encouraged to read \cite{CHK00}.



The above correlations between CRF variability and climate
(and in particular, cloud cover), indicate that CRF modulations
appear to be responsible for climate variability, most probably through modulation of the 
amount of LACC.
Nevertheless, since
all of the above CRF variability ultimately originates from solar
activity changes, it is not possible to unequivocally rule out
the possibility that the CRF/climate correlations are
coincidental, and that both are independently modulated by
solar activity with similar lags.

An independent CRF/climate correlation on a much longer time scale, in which variations in the CRF do not originate from solar variability, was found
by \cite{Shaviv02a,Shaviv02b,Shaviv03}. It was shown using astronomical data 
that a large $\sim{\cal O}(1)$ CRF variability should arise from our
passages through the galactic spiral arms, with a period of $132
\pm 25$~Ma. It was also shown that the
CRF history can actually be reconstructed using the cosmic-ray exposure age
data of Iron meteorites, exhibiting a periodicity of $143\pm 10$~Ma 
and a phase consistent with the astronomical
data. Moreover, it was found that the reconstructed CRF nicely synchronizes
 to the occurrence of ice-age epochs on Earth, which
appeared on average  every  $145\pm 7$~Ma over the past billion years. Additionally, 
the mid-point of the ice-age epochs is predicted to lag by $31 \pm 8$~Ma
after the mid-point of the spiral arm crossing, while it is
observed to lag by $33 \pm 20$~Ma. That is, the CRF and ice-age
epoch signals agree in both phase and period. The same analysis also
revealed that the long term star-formation activity of the Milky Way correlates with
long term glacial activity on Earth. In particular, a dearth in
star formation between 1 and 2 Ga before present, coincides with
a long period during which glaciations appear to have been
totally absent \citep{Shaviv02b,FaintSun}.

We should also point out several experimental results supporting, though not proving yet,  a CRF/cloud cover link. 
\cite{Harrison} found experimentally that CN
formation is correlated with natural Poisson variability  in cosmic ray showers. In
other words, this link appears to be more than hypothetical.
In another set of experiments, it was shown that cosmic rays play a decisive role in the formation of small clusters \citep{Arnolds}. If these small clusters can be shown to grow quickly enough, as opposed to being scavenged by large particles, the link between cosmic rays and the formation of cloud condensation nuclei and ultimately cloud cover could be firmly established. 

We will not dwell here on the actual mechanism responsible for CRF link with cloud behavior.  We will simply {\em assume} henceforth that this link exists, as supported by empirical and experimental data, even though it is still an issue of debate. This point has to be kept in mind since the conclusions we shall reach, will only be valid if this assumption is correct. 

Using the above assumption, we study several time scales to see whether estimates on global temperature sensitivity can be placed, together with estimates on the CRF/temperature relation. We will do so by comparing the observed temperature changes with changes in the radiative budget, an approach previously pursued in numerous analyses (e.g., \cite{Hoffert92,Hoffert96,Hansen93,Gregory02}). This method for obtaining the global temperature sensitivity using paleodata is orthogonal to the usage of global circulation models (GCMs) upon which often quoted results are based \citep{IPCC}. Hence, the two methods suffer from altogether different errors. It is therefore clearly advantageous to follow this path as an independent estimate.  For example, \cite{Cess} have shown that the large uncertainty in the sensitivity obtained in GCMs stems from the uncertain feedback of cloud cover. Since we use the actual global data, all the feedbacks are implicitly considered. The main contribution in this work is to specifically consider the contribution of the CRF to the changed radiative budget. As a note of caution, one should keep in mind that the most notable assumption in this method is the quantification of climate sensitivity with one number. In other words, it assumes that on average Earth's climate responds the same irrespective of the geographic, temporal or frequency space distribution of the radiation budget changes. It also assumes that different radiative forcings act linearly. 

Once the radiative forcing and temperature changes are obtained, the sensitivities can be estimated with
\begin{equation}
\label{eq:baseeq}
 \lambda \equiv \left.d T_{global}\over d F\right|_{F=F_{0}} \approx {\Delta T \over \Delta F}.
\end{equation} 
$\Delta F$, which is the globally averaged change in the radiation flux (per unit surface area), will also include here the contribution $\Delta F_{CRF}$ arising from a changed energy flux $\Phi_{CR}$ of cosmic rays. Note also that over short time scales, $\Delta T$ or $\Delta F$ have to be properly modified to include the finite heat capacity of the system, and the consequent finite adjustment time it has.  We should also consider the possibility that $\lambda$ is dependent on the temperature. For example, the positive climate feedback arising from the formation of ice sheets could increase the sensitivity of a glaciated Earth, while the reduced atmospheric water vapor content, can reduce the sensitivity. 

In addition to $\lambda$, we will also estimate the sensitivity to CRF variations, defined as:
\begin{equation}
\mu \equiv -\Phi_0 \left. dT_{global} \over d \Phi_{CR} \right|_{\Phi_{CR}=\Phi_0} ,
\end{equation}
where $\Phi_{CR}$ is the cosmic ray energy flux reaching Earth at energies of $\sim$~10~GeV (the energies responsible for tropospheric ionization), while $\Phi_0$ is the average flux reaching today. 

\section{Radiative forcing of low altitude cloud cover}
\label{sec:clouds}

Without a detailed physical model for the effects of cosmic rays on clouds or a detailed enough record of radiation budget measurements correlated with the solar cycle, it is hard to accurately determine the quantitative link between CRF variations and changes in the global radiation budget. In particular, it is hard to do so without limiting ourselves to various approximations.
Nevertheless, this link is important since it will be used in most of our estimates for the global temperature sensitivity.

The basic observation we use to estimate the radiative forcing of clouds is the apparent correlation between CRF variations and the {\em amount} of low altitude cloud cover. A naive approximation is to assume that the {\em whole} climatic effect can be described by variations in the extent of the cloud cover, namely, that we neglect effects in the cloud properties, or possible climatic effects associated with atmospheric ionization but not with clouds. It also implies that the geographical distribution of the effect is the same as low altitude clouds on average. We will first estimate this ``zeroth" order term and then try to estimate the possible contribution of other corrections.

\underline{Amount of cloud cover:}
Over the solar cycle, the varying CRF appears to cause a 1.2 to 2.0\% (absolute) change in the amount of LACC  \citep{Marsh00,Kirkby00,Marsden03,CHK00}. We will therefore adopt a change of $1.6 \pm 0.4 \%$ in the LACC. 

The total radiative forcing of the LACC is estimated to be $-16.7~\Wm$  from the average 26.6\% cloud cover \citep{Hartmann1992}. If one however compares the forcing of the total cloud cover from different hemispheres and different experiments (Nimbus and ERBE, \cite{Ardanuy91}), one finds variations which are typically $2.5 \Wm$ on the $\sim 50 \Wm$ shortwave (SW) ``cooling" and $7\Wm$  on the $27\Wm$ longwave (LW) ``warming". Since LACC typically comprise {\em half} of the total amount cloud cover, an error of $\sim 4\Wm$ is to be expected.

Thus, the changed radiative forcing $\Delta F_f$ associated with the varying amount of cloud cover, should be $-1.0 \pm 0.35 \Wm$. This implicitly  {\em assumes} that the incremental cloud cover has the same average net radiative properties as the whole 27\% of the LACC.

\underline{Cloud Optical Depth:}
 Changes in the cloud properties could take place {\em in addition} to changes in the cloud amount. According to \cite{Marsden03}, there is a small negative correlation between the average LACC opacity $\bar{\tau}$ and the varying CRF. Over the solar cycle, $\bar{\tau}$ changes by $-4\%$ relative to its global average of about 4 in regions defined to be covered by LACC. 

Is such a change in $\bar{\tau}$ reasonable? According \cite{Marsden03}, there are two limiting cases for the effects on cloud properties. The first is changing the number density of cloud condensation nuclei (CCN) given a fixed amount of Liquid Water Content (LWC), that is, CCN limited. This is similar to the ``Twomey  effect" where enhanced aerosol density affects the droplet size and cloud albedo \citep{Twomey1977,Rosenfeld2000}. The second case is increasing the CCN density together with the LWC, and obtaining similar sized drops (LWC limited). Although the two cases are plausible, they do not change the cloud properties in the same way.

One can show that a cloud's optical depth for SW absorption is (e.g., \cite{Marsden03}) 
\begin{equation}
\tau \approx {3 \over 2} {\rho_\mathrm{eff} \Delta z \over \rho_0 R_\mathrm{eff}},
\end{equation}
where $\rho_0$ is the density of water, $\rho_\mathrm{eff}$ is the mass loading of water (i.e., its effective density), $\Delta z$ is the vertical extent of the cloud and $R_\mathrm{eff}$ is the effective radius of the cloud droplets, defined as the ratio between the 3$^{\mathrm rd}$ and 2$^{\mathrm nd}$ moments  of the droplet distribution ($\left<r^3\right>/\left<r^2\right>$).  In the case of a CCN limited condensation, $R_\mathrm{eff} \propto n_{CCN}^{-1/3}$, and $\tau$ will increase with $n_{CCN}$, while in the LWC limiting case, $\rho_\mathrm{eff} \propto n_{CCN}$ and $\tau$ will increase as well. One can therefore write:
\begin{equation}
{\delta{\tau} \over \tau} =  \beta {\delta n_{CCN} \over n_{CCN}},
\end{equation}
with $\beta = 1$ for LWC limited case and $\beta=1/3$ for the CCN limited case. Since the lower troposphere ionization rate changes by about 7\% between solar minimum and maximum, we should expect to get  at most a similar increase in the CCN. Thus we should expect ${\delta{\tau} / \tau} \lesssim 7\%$. The fact that a negative change in $\tau$ was observed \citep{Marsden03}, could arise because the increase in cloud lifetime results with thinner clouds on average.

Next, one can approximate the relation between $\tau$ and cloud albedo ${\cal A}$, by the relation \citep{Hobbs}:
\begin{equation}
 {\cal A} \approx { \tau \over \tau + \tau_{1/2}},
\end{equation}
where $\tau_{1/2} \approx 6.7$ for an asymmetry parameter of 0.85 \citep{Hobbs}. Once we differentiate, we find:
\begin{equation}
 {d {\cal A} \over d\tau} \approx {{\cal A}^2 \tau_{1/2} \over \tau^2}.
\end{equation} 
If we consider the transmission ${\cal T} \approx 0.75$ of the atmosphere (to obtain a top-of-atmosphere albedo, from a cloud-top albedo), that the LACC covers only a fraction $f_{low}$ of the globe, and that the average top-of-atmosphere incidence of radiation is ${\bar F} = 344 \Wm$, we find that the change in albedo is responsible for a changed radiation budget of:
\begin{equation}
  \Delta F_{{\cal A}} \approx {{\cal A}^2 \tau_{1/2} \over \tau^2} {\bar F} {\cal T}^2 f_{low} \delta {\bar{\tau}} \approx + 0.13 \Wm.
\end{equation} 
The positive sign implies that the small apparent reduction in ${\bar{\tau}}$ contributes a small warming contribution.

If we had no knowledge of $\bar{\tau}$, changes in it could have resulted with a correction to $\Delta F_{\cal A}$ which are only as large as $-0.23 \Wm$ (for the LWC limited case, and $\delta \tau/\tau \lesssim 7\%$). We take this uncertainly in ${\bar{\tau}}$ as another source of error for the radiative forcing $\Delta F$.

\underline{Cloud Emissivity:}
There could still be more physical terms contributing to $\Delta F$. If the LWC in the clouds can vary as well (that is, the clouds are not CCN limited but rather water limited), then also the IR emissivity can change. It will do so by changing the emissivity, relative to black body (e.g., \cite{Stephens78}) which is given by:
\begin{equation}
\epsilon(\delta z) = 1- \exp(- \tau_{IR}).
\end{equation}
where we have defined $\tau_{IR} \equiv a_0 \rho_\mathrm{eff} \delta z$. Here, $\rho_\mathrm{eff}$ is the liquid water content,  $a_0$ is the mass absorption coefficient (for water clouds, $a_0 \approx 0.13 m^2g^{-1}$ \citep{Stephens78}) and $\delta z$ is the width of the cloud layer above a given point.
By changing the emissivity, we change the outgoing long-wavelength flux by 
\begin{equation}
\Delta F_{IR} = {\cal T} \sigma T^4 f_{low} \Delta \epsilon = \exp( -\tau_{IR}) {\cal T} \sigma T^4 \tau_{IR}{\Delta \rho_\mathrm{eff} \over \rho_\mathrm{eff} }
\end{equation}
where ${\cal T} \sim 0.6$ is the transmittance of the atmosphere to IR, above the cloud. 
For typically small values of $\rho_\mathrm{eff}$ of $0.3 g/m^3$, and $\delta z = 100 m$ (which would give the largest effect), we get corrections of $\Delta F_{IR} = 1.1 \Wm (\Delta \rho_\mathrm{eff} / \rho_\mathrm{eff}) \lesssim 0.1 \Wm$. This positive flux outwards tends to cool (i.e., increase the CRF/temperature effect), but it is a small effect. 

By changing the emissivity, we can also shift the apparent location of the top of the clouds, and with it their temperature. In other words, we should expect outgoing LW radiation to come from higher up the atmosphere where the temperature is lower.

A higher $\rho_\mathrm{eff}$ will shift the IR emission ``surface" vertically by typically:
\begin{equation}
 \Delta z \sim {1\over a_0 \rho_\mathrm{eff}}\left({\Delta\rho_\mathrm{eff}\over \rho_\mathrm{eff}}\right).
\end{equation}
Using the black body law, the change in the radiative emission over the solar cycle will therefore be less than
\begin{equation}
-\Delta F_{\Delta T} \lesssim f_{low} 4  \sigma T^3 \Delta z {d T \over dz} \lesssim 0.2 {\Delta\rho_\mathrm{eff} \over\rho_\mathrm{eff}} \Wm,
\end{equation}
once globally averaged. For the last inequality, we took a typically low $\rho_\mathrm{eff}$ of $0.3 g/m^3$, a wet adiabat of $dT/dz \sim 0.6^\circ K (100 m)^{-1}$ and $f_{low} \approx 0.28$. Since $\Delta \rho_\mathrm{eff}/\rho_\mathrm{eff}\lesssim 0.1$, this effect will be even smaller at best (and in opposite sign as the previous effect).

This result is also reasonable considering that the {\em total} long wavelength heating effect of LACC was estimated to be $\sim 3.5~\Wm$ \citep{Hartmann1992}, while cloud albedo is responsible for a globally averaged cooling of $\sim 20~\Wm$, implying that changes in albedo will likely be more  important for changing the radiative budget arising from LACC variations.

\underline{Ocean Bias:} Additional unaccounted effects are possible. For example, as we have previously stated, we implicitly assume that the incremental change in the LACC contributes to the radiative budget as the LACC on average. This need not be the case. Suppose the CRF-LACC effect primarily takes place over the oceans. This is reasonable because it is the oceans where the density of CCN is lowest. Since the oceans have a lower albedo than land, covering ocean surface has a larger net decrease in the radiative budget than covering land mass. 

If the land albedo is typically higher by 15\% and LACC is typically half as frequent over land, then the albedo change arising from uncovering clouds over only oceans is higher than the albedo change obtained when uncovering the average LACC, by $\Delta {\cal A} \sim 3\%$. This corresponds to a flux change of $\Delta F_{\Delta {\cal A}} \approx \Delta {\cal A} {\cal T}^2 \Delta f_{low} {\bar F} \approx + 0.1 \Wm$. That is, a likely ocean bias implies that we are slightly underestimating $\Delta F_{CRF}$.

\underline{Other effects:} If the effect is geographically localized to certain areas, then a larger discrepancy could arise if the radiative properties of the LACC over those geographic regions is significantly different from the properties of LACC on  average. A correlation map between LACC variations and CRF change \citep{Marsh00}, reveals that some regions (particularly over oceans) stand out with a higher correlation than others. Nevertheless, they do not appear to cluster around particular latitudes or other special regions. Thus, the assumption of geographic uniformity may be not that bad.

Another hard to estimate effect could arise from the expected increase in cloud lifetime. For example, cumulus-type clouds could penetrate into higher altitudes, thereby reducing their IR emission. 

Thus, until we fully understand all the details in the physical picture, we should take the estimated radiative forcing and the error with a grain of salt. Taking the above into considerations, our best estimate for the radiative forcing of the cloud cover variations over the solar cycle is $\Delta F_{CRF} = 
\Delta F_f + \Delta F_{\cal A} + \Delta F_{\Delta {\cal A}} = 
1.0 \pm 0.4 \Wm$, globally averaged (we have neglected $\Delta F_{IR} \lesssim 0.1 \Wm$ and $-\Delta F_{\Delta T} \lesssim 0.02 \Wm$).
This should be compared with the $0.1\%$ solar flux variations, giving rise to an extra ``direct" forcing of $\Delta F_{\mathrm{flux}} \approx 0.35 \Wm$ \citep{Frohlich}.

A related number describes the relation between CRF variations and changes in the low altitude ionization rate. At energies which can reach  the lower troposphere and low geomagnetic latitudes ($\gtrsim 10 GeV$), a fair representative of the flux are the neutron measurements at Hunacayo Peru and Halaekala, Hawaii (near the geomagnetic equator, with a rigidity cutoff of 12.9 GeV). There, the amplitude in the CRF variations over the past few solar cycles was $7 \pm 2$\% \citep[e.g.,][]{Sven1998}
Thus, we find that the radiative sensitivity to CRF variations is about 
\begin{equation}
\alpha  \equiv -\Phi_0 {d F \over d \Phi_{CR}} = 14 \pm 6 \Wm.
\end{equation}

 \section{Estimating Climate Sensitivity}
 
We now proceed to estimate the climate sensitivity. We do so by comparing the radiative forcing change between two eras to the temperature change which ensued, using eq.~\ref{eq:baseeq}. In most estimates, we will rely on the results of \S\ref{sec:clouds} to obtained the contribution of the changed CRF to the changed radiative forcing. These include seven different comparisons, spanning from variations over the solar cycles, to variations over the Phanerozoic as a whole. Subsequently, we will combine the results to obtain our best estimate for the climate sensitivity.
 
\subsection{The $\Delta T$/CO$_2$ correlation over the Phanerozoic}
\label{sec:PhanCO2}

\cite{Shaviv03} have shown that more than 2/3's of the variance in the reconstructed tropical temperature variability $\Delta T_{trop}$ over the Phanerozoic can be explained using the variable CRF, which could be reconstructed using Iron meteorites. On the other hand, it was shown that the reconstructed atmospheric CO$_2$ variations do not appear to have any clear correlation with the reconstructed temperature.  The large correlation between reconstructed CRF and temperature is seen in fig.~\ref{fig:Phanerozoic}. It is this correlation which led the authors to conclude that the Phanerozoic climate is primarily driven by a celestial driver. The lack of any apparent correlation with CO$_2$ was used to place a limit on the global climate sensitivity. 

A subsequent analysis by \cite{Royer} has shown that pH corrections could have been important at offsetting the $\delta^{18}$O record upon which the temperature reconstruction is based.
In particular, The pH correction term of \cite{Royer} has the form:
\begin{equation}
\Delta T_{pH} = a \{\log R_{CO2} + \log \Lambda(t) - \log \Omega (t) \}, 
\end{equation}
where $R_{CO2}$ is the atmospheric partial pressure of  relative to today, $\Lambda(t)$ is $(Ca)(t)/(Ca)(0)$ - the mean concentration of dissolved calcium in the water relative to today, while $\Omega (t)$ is $[Ca^{++}][CO_{3}^{--}]/K_{sp}$ at time $t$ relative to today. The value of $a$ obtained in \cite{Royer} assumes no ice-volume correction, yielding $a = 3.4\deg$. Once the ice-volume effect on $\delta^{18}$O is considered \citep{Veizer2000,RoyerReply}, one obtains: $a\approx  1.4\deg$. 

Since the pH correction depends on $R_{CO2}$, so will the corrected temperature. A simple correlation between the corrected temperature and the reconstructed $R_{CO2}$ will then be meaningless. Instead, the method to proceed is to defined a CO$_2$ ``uncorrected" temperature as:
\begin{equation}  
\Delta T' = \Delta T - a \log R_{CO2}. 
\end{equation}  
Any correlation that this signal will have with $R_{CO2}$ will then be real, since this ``temperature" depends only on $\delta^{18}$O, and the small $\Lambda$ and $\Omega$ terms.  This uncorrected temperature can then be fitted with
\begin{eqnarray}  
\Delta T'_\mathrm{model} &=& A + B t + (C - a \log_{10}2) \log_{2} R_{CO2}  \\ & & + D g(\Phi_{CRF}(t)). \nonumber  
\label{eq:Ttag}
\end{eqnarray}
$A$ and $B$ allow for systematic secular trends in the data. $D$ relates the cosmic ray energy flux $\Phi(t)$ to $\Delta T$ \citep{Shaviv03}. The term $(a \log_{10} 2) log_2 R_{CO2}$ was added such that $C$ will keep its original meaning in \cite{Shaviv03}, which is the tropical temperature increase associated with a doubled $R_{CO2}$.

The lack of a correlation between $\delta^{18}$O and $R_{CO2}$ \citep{Shaviv03} originates from the fact that $(C - a \log_{10}2)$ happens to be coincidentally close to 0. In other words, the pH correction to $\delta^{18}$O  and $\Delta T$, happens to be similar to the tropical temperature sensitivity to changes in $R_{CO2}$ (Without the pH correction, the preferred value for $C$ in the absence of correlation is not $a \log_{10}2$, but 0). Scientifically, this is somewhat unfortunate, because without this coincidence the $\delta^{18}$O  signal would have had a clear correlation with the $R_{CO2}$ signal, and the $R_{CO2}$ fingerprint would have been discernible in the Phanerozoic data. 
  
If we repeat the analysis of Shaviv and Veizer (2003), and consider also the effects of $a {\log \Lambda(t) - \log \Omega(t)}$ introduced by \cite{Royer}, and corrected for $R_{CO2}$ as described above, we obtain:  $C = 0.69\deg$ (or an upper limit of 1.12, 1.42 and 1.73$\deg$ at 68\%, 90\% and 99\% confidence levels, and a lower limit of 0.39, 0.10, -0.21$\deg$, respectively). 
This gives $\lambda = 0.28 \pm 0.15 \CWm$. 

Without the effect of cosmic rays (i.e., with the $D$ term removed in the model given by eq.\ \ref{eq:Ttag}), more of the reconstructed temperature variability can be explained with CO$_2$, and the estimate range for $\lambda$ broadens respectively to $\lambda = 0.36 \pm 0.22 \CWm$. More limits are given in table 1. 

Note that this estimate is independent of  $\alpha$ determined in \S\ref{sec:clouds}. The first range for $\lambda$ simply assumes that a CRF/climate link exists, while the second quoted range, even neglects this assumption.

\subsection{CRF/$\Delta T$ correlation over the Phanerozoic}
\label{sec:PhanTotal}

The significant correlation between CRF and temperature over the Phanerozoic  was also used to place limits on the ratio between CRF variations and temperature change.  Together with the results of \S\ref{sec:clouds} we can place a limit on $\lambda$.

In \cite{Shaviv03}, it was found that if $\Delta T_{trop}$ is approximated with $\Delta T_{trop} = D \left[(\Phi_{CR}/\Phi_0)^p-1\right]$, where $p=1/2$ and $\Phi_0$ is the CRF today, then $D = 8 \pm 4 \deg$.  Almost all the contribution to the error arises from our limited knowledge of the actual amplitude of the CRF variations. If we generalize to other power laws $p$ between 0.25 to 1.5, and repeat the procedure described in \cite{Shaviv03}, we find that 
\begin{equation}
\hskip -5pt \mu \equiv \left. - \Phi_0 {d  T_{global}  \over d \Phi_{CR}} \right|_{\Phi_{CR} = \Phi_0}\hskip -15pt = p D {\delta T_{global} \over \delta T_{trop}} = 6.0 \pm 3.0 \deg,
\end{equation}
where we have taken $\delta T_{global} / \delta T_{trop} \approx 1.5$ as typically obtained in GCMs \citep{IPCC}. The reason the error does not increase much once we introduce a range of $p$'s is because $\mu$ (but not $D$) is rather insensitive to $p$. Note also that $p=1/2$ is theoretically preferred \citep{Yu,Harrison,Ermakov}, and a small $p$ is also favored by the empirical data (see fig.~\ref{fig:Phanerozoic}).

Using the result for $\alpha$ obtained in \S\ref{sec:clouds}, we find
$ \lambda = {\mu / \alpha} = {0.44^{+0.45}_{-0.22}} \CWm $.
where we quote the median $\lambda$ and the 16$^{\mathrm{th}}$ and 84$^{\mathrm{th}}$ percentiles (1-$\sigma$). More details on the distribution appear in fig.~\ref{fig:pdfs1} and table 1.


\figureone

\subsection{Bounds from the total $T$ and $\Phi_{CR}$  variations over the Phanerozoic}
\label{sec:PhanAlbedo}

Using the same Phanerozoic data and an altogether different set of argumentations, we can place additional limits on $\lambda$ and on $\mu$.
We do not know accurately
  how large are the absolute CRF variations that give rise to the temperature oscillation over the Phanerozoic (hence the relatively large error on $\mu$ and $D$). Nevertheless,  we know that there is a maximum increase of $\sim  2\deg$ in the tropical temperature above today's tropical temperature, once averaged over the 50 Ma time scale \citep{Veizer2000}. This  approximately corresponds to an increase of $\Delta T \sim 3\deg$ globally. We assume that this could arise by removing at most  $f_{min}\sim 80\%$ of the LACC, which would give rise to a global cooling of $f_{min} \Delta F_{LACC} \approx f_{min} 17 \Wm$. Namely, 
\begin{equation}
  \lambda_{min} \gtrsim  {\Delta T \over f_{min} \Delta F_{LACC}} \approx 0.22 \CWm.
\end{equation}
This is an absolute minimum for the climate sensitivity. Otherwise, the CRF-temperature link observed over the Phanerozoic would require too large a radiative budget change to be explained by LACC variations. 

A similar limit can be placed on $\mu$. The largest increase over the Phanerozoic  of $\Delta T \sim 3\deg$ relative to today,  and which can be attributed to CRF variation, cannot arise from a flux decrease larger than 85\%, since larger reductions are inconsistent with various astrophysical constraints. This implies $\mu \gtrsim 3.0 p/(1-0.15^p) \deg $, if a relation of the form $\Delta T \propto \Phi^p$ exists. Likewise, the maximum temperature decrease over the Phanerozoic is about 2.5$\deg$ over the tropics, or $\sim 3.75\deg$ globally, while the astronomical constraints give a maximum flux increase of about 95\% relative to today for models still satisfying astronomical constraints \citep{Shaviv02b}. This gives $\mu \gtrsim 3.75 p / (1.95^p - 1) \deg $. Combining the two, we get $\mu \gtrsim 3.7\deg $ if $p$ is unconstrained, or $\mu \gtrsim 4.7 (\hskip 2pt \mathrm{or} \hskip 2pt 4.0) \deg$ if $p=1/2 (\hskip 2pt \mathrm{or}\hskip 2pt 1)$.

Using \S\ref{sec:PhanTotal}-\ref{sec:PhanAlbedo}, we thus obtain $\mu = 6.5 \pm 2.5 \deg$.

\subsection{Cretaceous and Eocene Climates}

Particular geological epochs were studied under more scrutiny, and without being averaged out on the 50 Ma time scale, as the Phanerozoic data was. In particular, there were estimates for both the radiative forcing and temperature change of two geological periods which were particularly warm relative to today. One is the Cretaceous, at about 100~Ma before present, and the second is the Eocene, at 50~Ma.

\cite{Barron93} estimated that the mid-Cretaceous was $7\pm 2\deg$ warmer than today, and that it arose from an increase of $8 \pm 3.5 \Wm$ in the radiative budget. However, \cite{Hoffert96} point out that this estimate included only the change from the increased amount of atmospheric CO$_2$ and it did not include the increased forcing associated with surface albedo changes. Once taken into account, the \cite{Hoffert92} estimate for the radiative forcing increases to $15.7 \pm 6.8 \Wm$. Their estimate for the temperature increase is also larger at $9 \pm 2\deg$. We adopt the \cite{Hoffert92} estimate here.

\cite{Hoffert96} estimated the temperature and radiative flux increases associated with the Eocene. They are $4.3 \pm 2.1 \deg$ and $11.8 \pm 3.6 \Wm$ respectively. The temperature and forcings can be used to estimate the sensitivity through $\lambda = \Delta T/\Delta F$. The results for the Eocene and Cretaceous are $ \lambda = {0.37^{+0.20}_{-0.16}} \CWm $ and $ \lambda = {0.57^{+0.44}_{-0.18}} \CWm $ respectively. They are also summarized in fig.~\ref{fig:pdfs0} and table 1.

These estimates do not include however the possibility that CRF variations affect climate. To estimate this effect, we estimate the CRF differences between the two geological periods and today using the CRF reconstruction described in \cite{Shaviv02b}. We then calculate $\Delta F_{CR}$ arising from the CRF change and use $\Delta T = \lambda (\Delta F_0 + \Delta F_{CR})$ to obtain $\lambda$.

We find that the CRF was 20\% to 70\% of the flux today during the mid-Cretaceous. Through the effects on clouds, this should have contributed towards an increase in temperature, and therefore reduce our estimate for $\lambda$.  During the Eocene, $\Phi_{CR}$ should have been between 0\% and 20\% higher than today. From the 6 epochs described here, this is the only case in which the effect of the CRF is to increase the estimate for the climate sensitivity.

The radiative forcing associated with the CRF variations can be estimated using the value of  $\alpha$. Numerically, we find $\Delta F_{CRF} = (-10 \pm 10\%) \alpha =-1.4 \pm 1.5 \Wm$ for the Eocene and  $\Delta F_{CRF} = (70\pm40\%) \alpha = 10.0 \pm 6.0 \Wm$ for the Cretaceous (if we limit ourselves to $p$ between 1/2 and 1). The new estimates  are $ \lambda = {0.36^{+0.24}_{-0.11}} \CWm $ for the Cretaceous, and $ \lambda = {0.42^{+0.25}_{-0.18}} \CWm $ for the Eocene. (More detail at fig.~\ref{fig:pdfs1} and table 1).

\figuretwo

\subsection{Warming since the last glacial maximum}
\label{sec:LGM}

Several studies have attempted to estimate the global sensitivity by comparing the temperature increase since the last glacial maximum (LGM) with the radiative forcing change. For example, \cite{Hoffert92} estimate that a radiative forcing of $\Delta F = 6.7 \pm 0.9  \Wm$ is reponsible for a temperature increase of $\Delta T = 3  \pm 0.6 \deg$. On the other hand, \cite{Hansen93} find a higher sensitivity. This is because they find that a similar radiative forcing of $\Delta F = 7.1 \pm 1.5  \Wm$ is responsible for a much larger $\Delta T = 5 \pm 1\deg$ increase. The main difference is that Hoffert \& Covey base their temperature esimate on the oceanic CLIMAP temperature reconstruction, while Hansen et al. based theirs on land temperature proxies. We will adopt the average temperature change and increase the error to be conservative. Namely, we choose $\Delta T = 4 \pm 1.5 \deg$. Similarly we take $\Delta F = 6.9 \pm 1.5  \Wm$. This gives $ \lambda = {0.58^{+0.29}_{-0.20}} \CWm $  (as detailed in fig.\hskip 2pt\ref{fig:pdfs0} and table 1).

Again, the above estimates do not include the net radiative forcing change due to CRF modulation of the cloud cover. 

\cite{Christl2003} and \cite{Frank1997} assumed that $^{10}$Be flux modulation on this time scale is primarily a result of modulation by the varying geomagnetic field. Using this flux they derived that the geomagnetic field was about 50\% its present value at 20 ka before present.  Moreover, the total effect of the terrestrial field is a $\sim 20\%$ reduction of the CRF penetrating down to $\sim 3~km$, where low altitude clouds typically form \citep{Compton}. Namely, the 50\% reduction in the global magnetic field should correspond to a reduction of about 10\% in the relevant CRF. i.e., $\Delta F_{CRF} \approx 0.1 \alpha $. This should be compared with the 7\% variation over the solar cycle. 

\cite{Sharma} relaxed the assumption that the $^{10}$Be flux modulation is predominantly terrestrial. By using independent proxies for the terrestrial field, he obtained that the field was only 30\% lower than today, corresponding to a $\sim 6\%$ reduction in the high energy CRF.  The rest of the  $^{10}$Be flux variations, were attributed to a reduced solar modulation factor $\phi$ \citep{Masarik1999}, that at 20 ka was about a 1/3 of its current day average value of $\sim$550 MeV. If we take the expression for the CR differential number flux  $J(E_p,\phi)$ given in \cite{Masarik1999} and integrate $\int_{E_0}^\infty E_p J(E_p,\phi) dE_p$ to get the total flux reaching the troposphere (i.e., with $E_p \gtrsim E_0 \approx 10 GeV$), we find that the increased solar activity as borne out from the increased solar modulation factor was responsible for an additional 6\% reduction in the high energy CRF. We therefore get a total reduction of about 12\% in the CRF, and $\Delta F_{CRF} \approx 0.12 \alpha $. We take this value.

Since we find that a larger total radiative forcing is responsible for the same temperature change, we obtain a lower estimate, $ \lambda = {0.48^{+0.22}_{-0.16}} \CWm $  (also detailed in fig.~\ref{fig:pdfs1} and table 1).

Instead of using the radiative forcing through cloud cover modification, we can use our limits of $\mu$ which are not based on LACC forcing, but instead on the observed temperature change over the phanerozoic. We found $  \mu = 6.5 \pm 2.5 \deg$. Thus, the 12\% decrease in the CRF causing low altitude tropospheric ionization should translate into a warming contribution  of $\Delta T_{CRF} \approx  4.8$ to $10.8\deg$ warming. We now use an equation of the form $\Delta T = \lambda (\Delta F_0 + \Delta F_{CRF}) \equiv \lambda \Delta F_0 + \Delta T_{CRF}$, and obtain: $ \lambda = {0.47^{+0.27}_{-0.19}} \CWm $ . 
 
To summarize, the effect of the decreased CRF since the LGM is to reduce our estimate for $\lambda$ by about 20\%, which is smaller than the error in the estimate itself. This result is valid also if we don't believe the CRF-climate link is through LACC moduation, but merely that such a link exists.

\subsection{Warming over the past century}
\label{sec:century}

Climate sensitivity can also be estimated using the global warming observed over the past century once the radiative forcing with their uncertainties are estimated.

Since the time scale is relatively short, it is necessary to consider the finite heat capacity of  the oceans. We base our analysis here on the work of \cite{Gregory02} who tackled this problem by considering the heat flux into the ocean in the energy budget. The main difference between our modified analysis here and that of \cite{Gregory02}, is that we will also consider the radiative forcing associated with the decreased CRF over the past century. Unlike the warming since the LGM, where this was a small correction, here it is will prove to be a notable one.

Again, we assume that the CRF modulates the LACC and that its radiative forcing is given in \S\ref{sec:clouds}. 

\cite{Gregory02} compared the period 1850-1900 with 1950-1990. Since the CRF record does not go back far enough, we need to use proxy data. A good choice is the geomagnetic $aa$ activity index. The advantage of the $aa$ index is that it can directly compare the long secular trends to variations over the solar cycle. In particular, we find that the secular increase between the two periods is roughly the same as the total variation in the $aa$ index over the solar cycle. In both cases it is roughly 12nT.  If we further assume that the ratio between the secular CRF variations to those over the solar cycle is similar to the ratio in the $aa$ index, we find that the decrease in the $\sim 10$~GeV CRF between the above periods should be about 7\%, the typical min to max variations over the solar cycle. Using our estimate for the radiative forcing change in \S\ref{sec:clouds}, this corresponds to about $1.0 \pm 0.4\Wm$.

According to \cite{Gregory02}, 
\begin{equation}
 \lambda \overline{\Delta T} = {\overline F} - {\overline Q},
\end{equation}
where ${\overline F}$ is the change in the radiative forcing (i.e., in the energy balance), while ${\overline Q}$ is the average net heat flux which entered the oceans between the two periods.

In our modified case ${\overline F} = {\overline F}_0 + {\overline F}_{CRF}$, where ${\overline F}_0$ is the ``standard" radiative forcing that was estimated by \cite{Gregory02} to be ${\overline F}_0 = -0.3$ to $1\Wm$. It includes anthropogenic, volcanic, solar luminosity and aerosol contributions (with the last one contributing the largest uncertainty). ${\overline Q}$ was estimated to be $0.32 \pm 0.15 \Wm$ while $\overline{ \Delta T} = 0.335 \pm 0.033\deg$ (all at $2 \sigma$).

Like \cite{Gregory02}, we assume that the errors have a Gaussian distribution. Following their procedure, we calculate the probability distribution function (PDF) for $\lambda$. They are given in figs.~\ref{fig:pdfs0},\ref{fig:pdfs1}, for the CRF and no-CRF cases. The added complication in the CRF case is the extra PDF for $\alpha$, which implies that ${\overline F}_{CRF}$ has a PDF itself. The PDF obtained for the ${\overline F}_{CRF}=0$ case is the same as the result of \cite{Gregory02}.

Inspection of fig.\ \ref{fig:pdfs1} and table\ 1 reveals that $\lambda = {0.32^{+0.32}_{-0.10}} \CWm $  at 1-$\sigma$ confidence. This is a clear reduction from the results of \cite{Gregory02}, where the lower 16th percentile for $\lambda$ is $0.67 \CWm $ and there is no formal upper limit (This assumes a prior that $\lambda$ cannot be negative).
 
\figurethree

 \subsection{Variations over the solar cycle}
 \label{sec:solarcycle}
 
 The shortest time scale we study is that of the solar cycle. Since the expected signal arising from solar variability is small ($\sim 0.1\deg$) looking at the recent few cycles is problematic since internal variations (such as volcanic eruptions, ENSO and other oscillations as well as simple inter-annual variations, or ``internal noise") in the climate are large and can drown the solar signal. To overcome this problem, we will look at a much longer temperature record. In particular, we use the post-Little Ice age data (i.e., last 300 yrs) of the \cite{Jones98} thousand year long reconstruction of temperature for both hemispheres, which includes proxy data of tree rings, ice cores, corals, and historical documents. The catch is that the solar cycle is not stable, and the actual period varies between about 9 and 12 years. To overcome this problem, we do not perform a harmonic analysis. Instead, we fold and average the data over the varying solar cycle period. Namely, each year is asigned a  phase $\phi$ defined as the time since last sunspot minimum devided by the length of the particular cycle encompassing the given year. Once we do so, we can average all points within a phase bin and obtain the average temperature. Also, the internal variance in the temperature allows us to estimate the error in $\Delta T(\phi)$. The result is depicted in figure \ref{fig:solarcycle}.
 
Evident from the figure is the fact that the temperature has a near sinusoidal behavior. By performing a $\chi^2$ fit to a form $\Delta T(\phi) = {\overline{\Delta T}} + (a/2) \cos (2\pi (\phi-\phi_0))$ with $\phi_0$ the phase relative to the occurrence of maximum sunspot number, we find at the $1-\sigma$ confidence limit that:
 \begin{eqnarray}
 a &=& 0.09 \pm 0.03 \deg , \\
 \phi_0 &=& 1.0 \pm 0.6 \mathrm{\hskip2pt rad}.
 \end{eqnarray}
 The value of $\phi_0$ implies that the average temperature lags behind the maximum sunspot number by $1.8 \pm 1.0$ years. This is to be expected, because of Earth's finite heat capacity. As a result, the response to radiative perturbations is not only damped, but it is lagged as well.

Other analyses estimated the surface temperature variation over the 11-yr solar cycle. \cite{Douglass} found $a=0.11 \pm 0.02\deg$, while \cite{White} found $a=0.10 \pm 0.02\deg$. Together with the current result, we will adopt $a=0.10 \pm 0.02\deg$ for the temperature variation between solar minimum and solar maximum.
  
\figurefour

We now use the results of \S\ref{sec:clouds}. In particular, the above temperature variations are assumed to arise from the $\sim 1.6\%$ variations observed in the LACC, such that the forcing over the solar cycle is  $\Delta F_{LACC} =   1.0\pm 0.4 \Wm$. An additional contribution of $\Delta F_{flux} = 0.35 \Wm$ is due to changes in the solar flux. The sensitivity itself is then given by:
\begin{equation}
 \lambda = {\Delta T / d \over \Delta F}
\end{equation}
where $d$ is a damping factor which arises from the finite heat capacity of the climate system and its inability to reach equilibrium at a finite time. 

The value of the damping factor is not well known. In principle, it can be obtained in climate models, but these give a range of values. Using a simple Ocean/Climate model, \cite{Schlesinger} obtained a damping of about 0.25 on the 11-yr solar cycle time scale (and about 0.75 on the centennial time scale). Other more extensive simulations find that the 11-yr solar cycle is damped relative to the centennial scale by a factor of $\sim 0.54$ \citep{CUB97}, 0.33 \citep{RLH99} or by comparing solar forcing to actual climate responce, to $\sim 0.68$ \citep{Waple}. If we further consider that the centennial time scale is damped relative to the long term response, by a factor of $\sim 0.7-0.75$ \citep{Schlesinger,IPCC}, then the 4 estimates for the damping factor are encompassed within $d =0.35 \pm 0.15$, for periodic oscillations with an 11-yr period. 

Note that by resorting to GCMs for the estimate of the damping factor, we are somewhat unfaithful to the spirit of this work, which is to estimate the sensitivity independently to the usage of GCM simulations. Nevertheless, the damping we use is a characteristic describing the {\em relative} behavior of different time scales. We still avoid using the {\em absolute} sensitivity obtained in GCMs. Moreover, the analysis of \citep{Waple} does indicate that empirically, the damping factor is consistent with that obtained in GCMs. The fact  that this result is somewhat larger than GCMs on average, would imply that we maybe underestimating the damping factor, and with it, overestimating the climate sensitivity. 

For nominal values, we find that $ \lambda = {0.26^{+0.26}_{-0.11}} \CWm$. Without the effects of the CRF, a much larger sensitivity (of $ \lambda = {0.94^{+0.91}_{-0.35}} \CWm $ ) is obtained because the same temperature variations are then to be explained only by the relatively small solar flux variations.

\def\selfont{\fontsize{8}{10pt}\selectfont}
\begin{table*}
\selfont
\caption{  \small \sf
Limits on the sensitivity of global temperature to radiative forcing, using different methods while assuming that CRF {\em does} or {\em does not} affect climate. $\dTdbl$ is calulcated using a forcing of 3.71$\Wm$ \citep{Myhre}. The columns denote the values below which $\lambda$ (or $\dTdbl$) are expected to be found at the given probability, as obtained with the particular method, or, once the methods are combined. Either by assuming $\lambda$ is strictly constant, or by assuming it ca }
\begin{tabular}{  l     c   c   c  c  c   c   c   c  c c }
\hline \hline
Period  (Method) 
                        &  \multicolumn{5}{c}{$\lambda [\CWm$]}    & \multicolumn{5}{c}{$ \dTdbl [\deg]$}     \\
                        &  1\%     &   16\%    & 50\%     & 84\%      &   99\%   & 1\%     &   16\%   & 50\%    & 84\%  & 99\%       \\
\hline \hline
\multicolumn{11}{c} {Without the effect of Cosmic Rays} \\
\hline
Phanerozoic (CO$_2$/T)
                         &  0.02  &  0.19    &  0.36   & 0.63     &  1.05   & 0.1    & 0.5      & 1.3    & 2.3     &      3.9              \\
Cretaceous\tablenotemark{a}
                         &  0.29  &  0.38    &  0.57   & 1.01     &   *$^i$   & 1.1    & 1.4      & 2.1    & 3.8     &      *               \\
Eocene\tablenotemark{b}
                         & 0.03   &  0.21    &  0.37   & 0.56     &  0.87   & 0.1    & 0.8      & 1.4    &  2.1    &  3.3                \\
LGM\tablenotemark{b,c}\phantom{a}              
                        & 0.10    & 0.38     & 0.58    & 0.87     &  1.48   & 0.4    & 1.4      &  2.2   &  3.2    &  5.5               \\
20\th Century\tablenotemark{d} 
                        & 0.34   & 0.67     & 1.31   &     *          &       *       & 1.4     &  2.5       & 4.9    &   *         &  *                  \\
Solar Cycle
                        & 0.24  & 0.58      &  0.94 &     1.85     &       *        & 0.9    & 2.2      & 3.5    & 6.8     & *                   \\
\hline
Combined  ($\lambda=const$)
                        &  0.33  &    0.44  &  0.52   &  0.62    &  0.79        & 1.3    &   1.6    &  1.9    &  2.3   &  2.9        \\
Combined  ($\lambda=\lambda_0+b \Delta T$)
                        &  0.24  &    0.43  &  0.54   &  0.66    &  0.87        & 0.9    &   1.6    &  2.0    &  2.5   &  3.2        \\
\hline \hline
  \multicolumn{11}{c} {With the effect of Cosmic Rays} \\
\hline
Phanerozoic (CO$_2$/T)\tablenotemark{e} 
                      & 0.02 & 0.14 &  0.28  & 0.43           &   0.65     & 0.1 & 0.5 &   1.0 & 1.6   &  2.4         \\
Phanerozoic (CRF/T)\tablenotemark{f}\phantom{a}   
                       &    0.02 &    0.22 & 0.44     & 0.89     & *             & 0.1   & 0.8      & 1.6     & 3.3     &    *                 \\
Phanerozoic 
   (Clouds)\tablenotemark{f,j} 
                       &  0.22    &                &               &                 &                & 0.8    &               &              &             &                            \\
Cretaceous\tablenotemark{a,f}
                         &  0.13  &  0.24   & 0.36  & 0.60      &  1.94         &  0.5  & 0.9     &  1.3     &   2.2   &  7.2            \\
Eocene\tablenotemark{b,f,h}
                        & 0.03 & 0.24     & 0.42  &  0.67      &  1.32          & 0.1  &  0.9        &  1.6    &  2.5   &   4.9            \\
LGM  (LACC)\tablenotemark{b,c,f}
                        & 0.08  & 0.30      & 0.47   & 0.69     & 1.12          & 0.3    &  1.1     &  1.7   &  2.5    & 4.1             \\
LGM ($\mu$)\tablenotemark{b,c,g} 
                        & 0.04  & 0.28      & 0.47   & 0.74     & 1.30               & 0.2    &  1.0     &  1.8   &  2.8    & 4.8             \\
20$^\mathrm{th}$ Century\tablenotemark{f}  
                        & 0.16   & 0.22     & 0.32    & 0.64     & *                        & 0.6   & 0.8      & 1.2    & 2.4     &    *                \\
Solar Cycles & 0.08   & 0.15    & 0.26    & 0.52     & 2.30                     & 0.3   &  0.6      & 1.0    & 1.9     & 8.6             \\
\hline
Combined\tablenotemark{h}  ($\lambda=const$) 
                        &  0.17  &    0.24  &  0.29   &0.37 & 0.50  & 0.6    &   0.9    &  1.1    &  1.4   &  1.9      \\
Combined\tablenotemark{h}  ($\lambda=\lambda_0+b \Delta T$)
                        &  0.25  &    0.27  &  0.35   &0.45 & 0.68  & 0.9    &   1.0    &  1.3    &  1.7   &  2.5      \\
\hline \hline
\end{tabular}
\tablenotetext{a}{\selfont  Based on \cite{Hoffert92}.   
$^\mathrm{b}$  Based on \cite{Hoffert96}.     
$^\mathrm{c}$  Based on \cite{Hansen93}.}
\tablenotetext{d}{ \selfont Based on \cite{Gregory02}. 
$^\mathrm{e}$ Based on \cite{Shaviv03}.}
\tablenotetext{f}{ \selfont Assumes the CRF-climate is through modulaton of the LACC, with $\alpha = 14 \pm 6 \Wm$}
\tablenotetext{g}{\selfont Using $\Delta T_{CRF}$ and $\mu$ from the Phanerozoic data instead of $\Delta F_{CRF}$ and $\alpha$. }
\tablenotetext{h}{\selfont The combined PDF does not include the LGM ($\mu$) estimate or the Phanerozoic/Clouds bounds. }
\tablenotetext{i}{\selfont Limits for $\dTdbl$ larger than $10\deg$ are meaningless and therefore not quoted.}
\tablenotetext{j}{\selfont The lower limit obtained from the maximum cloud cover changes depends on systematic errors. The confidence limits are therefore meaningless.}
\end{table*}

\subsection{Combined Results}

We now proceed to combine the PDFs obtained in the cases described in fig.~\ref{fig:pdfs0}, when the CRF/climate link is neglected, and the cases described in fig.~\ref{fig:pdfs1}, when the CRF/LACC effect is included. We combine the results in two cases. In the first, we assume that the global temperature sensitivity is constant, namely, that it does not depend on the average terrestrial temperature. In the second case, we allow the sensitivity to be temperature dependent. 

\subsubsection{Constant Sensitivity}

When combining the results under the assumption that CRF does not introduce a radiative forcing, we can simply multiply the PDFs and renormalize the result. The reason is that the error in the estimates of all the $\Delta F_0$ and $\Delta T$ are presumably uncorrelated with each other, and also because we have no prior on the value of $\lambda$ (except perhaps that it should be positive). 

On the other hand, when combining the cases which include the CRF/LACC effect, we must bear in mind that some of the error arises from the uncertainty in $\alpha$ -- the relation between cloud cover changes and radiative forcing. This uncertainty enters 5 PDFs, and we cannot simply multiply them. To overcome this obstacle, we calculate the PDFs assuming a given $\alpha$. Then, the combined PDF is given by
\begin{equation}
 P_{all}(\lambda) = { \int \left[ \prod_{i=1}^6 P_i(\lambda,\alpha) \right] P_\alpha (\alpha) d\alpha \over \int \left[ \prod_{i=1}^6 P_i(\lambda,\alpha) \right] P_\alpha (\alpha) d\alpha d\lambda
}.
\end{equation}
Again, this assumes that we have no prior on $\lambda$, and that besides the dependence on $\alpha$, the PDFs are not in anyway correlated with each other.

In figs.\ \ref{fig:pdfscombined},\ref{fig:contours} and table 1, we plot and describe the combined PDFs obtained in the two cases. We find that  $\lambda = {0.52^{+0.10}_{-0.08}} \CWm $ if the CRF/Climate link is neglected, and that $\lambda = {0.29^{+0.08}_{-0.06}} \CWm $ if the CRF/LACC link is included. Values of upper and lower limits on $\lambda$ and $\dTdbl$ at different confidence limits are given in Table 1.

\subsubsection{Variable Sensitivity}

We now alleviate the assumption of a constant sensitivity and allow a linear relation in the form $\lambda = \lambda_0 + b \Delta T$, where $\lambda_0$ is the sensitivity today and $b = d\lambda/dT$. Because the errors are not Gaussian (the distributions are generally skewed towards higher $\lambda$'s) we cannot fit $\lambda(\Delta)$ using a simple linear least squares. Instead, we calculate

\begin{equation}
 \hskip -0.7cm P_{all}(\lambda_0,b) = { \int \left[ \prod_{i=1}^6 P_i(\lambda_0 + b \Delta T_i,\alpha) \right] P_\alpha (\alpha) d\alpha \over \int \left[ \prod_{i=1}^6 P_i(\lambda_0 + b \Delta T_i,\alpha) \right] P_\alpha (\alpha) d\alpha d\lambda db
},
\end{equation}
and $P_{all}(\lambda_0) = \int P_{all} (\lambda_0, b) db$. 

Here we find that  $\lambda = {0.54^{+0.12}_{-0.1}} \CWm $ if the CRF/Climate link is neglected, and that $\lambda = {0.35^{+0.10}_{-0.08}} \CWm $ if the CRF/LACC link is included. Values of upper and lower limits on $\lambda$ and $\dTdbl$ at different confidence limits are given in Table 1.

This is our best estimate for the global sensitivity. It translates into a CO$_2$ doubling temperature change of $\dTdbl = 1.3 \pm 0.4^\circ$K. With the CRF/climate effect neglected, this number is $\dTdbl = 2.0 \pm 0.5^\circ$K.
 
Another point worth mentioning is the fact that once the CRF/LACC climate link is included, the median values for $\lambda$ obtained using different periods differ from each other by typically 1-$\sigma$ or less, while without the CRF/climate considered, differences can be larger than 2-$\sigma$.  Namely, the CRF/climate effect markably improves the consistency of the data. This can be seen in fig.\ \ref{fig:results}.

\figurefive

\figuresix
\figureseven

\section{Discussion}

We compared the radiative forcing and temperature change over several different time scales, while taking into consideration the alleged link between CRF variations and temperature change. We found that the 6 different time scales can be used to place similar bound on the global climate sensitivity and, when possible, also on the quantitative relation between CRF variations and temperature changed.

Different time scales and methods suffer from different uncertainties, which have to be kept in mind. These include:
\begin{enumerate}
 \item Limits based on the geochemical record over the past 550 million years implicitly assume that estimates of temperature variations using $\delta^{18}$O is well known. In principle, various biases might distort this relation and produce a wrong temperature scale.
 
 \item Limits based on large temperature variations and radiative forcing, assume that one can deduce the sensitivity today, to small perturbations, from temperature changes associated with large  radiative forcings, or worse, when different conditions existed under which the sensitivity could have been different (e.g., over the past 550 milllion years ice sheets where mostly absent). 
 To address this point, we allowed for $\lambda$ to be a linear function of temperature, but found no significant dependence. 
 
 \item Most of the estimates of the climate sensitivity assume the CRF/climate link is through modification of the LACC and that the radiative forcing associated with it is known. This entails in it several assumptions: (a) That LACC changes observed by the ISCCP measurements, for example, are indeed well represented by changes in the {\em amount} of cloud cover as opposed to other cloud characteristics. (2) That the incremental cloud cover changes behave as the average. (3) That tropospheric ionization does not markably affect the global temperature through an effect not related to cloud cover modifications. These uncertainties will not be resolved without detailed understanding of how and to what extent does atmospheric ionization affects the formation of cloud condensation nuclei, which affect cloud cover.  Nevertheless, the value of the sensitivity could still be independently bracketed using argumentation which does not assume the relation is through the LACC variations.
\item Over the short time scales associated with the solar cycle, one of the main uncertainties is the damping factor in the effect of a changed radiative forcing. 
 \end{enumerate}
 
Having said that, the fact that about  half a dozen independent analyses based on paleoclimatic to recent data yield roughly the same sensitivity should indicate that we are probably not missing large radiative  
forcing terms. Otherwise, there is no reason, other than chance, to obtain results consistent with each other. Moreover, the notably improved agreement between the sensitivities obtained once the CRF/climate effect is taken into account, is yet another indicator that the effect is real (see 
panel A vs.\ B in fig.\ 5). 

Our best estimate is  $\dTdbl \approx 1.3 \deg$ ($0.9\deg<\dTdbl<2.5 \deg$ at the 99\% confidence level). This is at the lower end of the often quoted range of $\dTdbl=1.5$ to $4.5\deg$ \citep{IPCC} obtained from Global Circulation Models (GCMs). \cite{Cess} have shown that the climate sensitivity obtained in this type of simulations predominantly depends on how clouds are treated, and whether they contribute a positive or negative feedback. The models roughly give that $\lambda^{-1} \approx 2.2 Wm^{-2}/\deg - \Delta F_{cloud} / \Delta {\bar{T}}$ with $\Delta F_{cloud}$ being the feedback forcing of clouds associated with a temperature change of $\Delta {\bar{T}}$. Thus, for a GCM to be compatible with the results obtained here, a negative cloud feedback is required. One such example was suggested by \cite{LindzenIris}

On the other side of the coin, we can also rule out very small climate sensitivities. This can be used for example to place a limit on possible large negative feedbacks, or to a lower limit on the effect of anthropogenic greenhouse gas (GHG) warming. 

Since the beginning of the industrial era ($\sim 1750$), non-solar sources contributed a net forcing of $0.85 \pm 1.3 \Wm$ \citep{IPCC} (assuming the errors are Gaussian). Over the past century alone, this number is $0.5 \pm 1.3 \Wm$.  The main reason why the error is large is because of the uncertain ``indirect" contribution of aerosols, namely, their effect on cloud cover. It is currently estimated to be in the range $-1\pm 1 \Wm$ \citep{IPCC}. Thus, anthropogenic sources alone contributed to a warming of $0.13 \pm 0.33 \deg$ since the beginning of the 20$^{\mathrm{th}}$ Century.

The sensitivity result can also be used to estimate the solar contribution towards global warming. Over the past century, the increased solar activity has been responsible for a stronger solar wind and a lower CRF. Using results of \S\ref{sec:century}, the reduced ionization and LACC were responsible for an increased radiative forcing of $1.0 \pm 0.4\Wm$. In addition, the globally averaged solar luminosity increased by about $0.4\pm 0.1\Wm$ according to \cite{Solanki,Hoyt,Lean95}. Thus, increased solar activity is responsible for a total increase of $1.4 \pm 0.4 \Wm$. Using our estimate for $\lambda$, we find $\Delta T_{solar} = 0.37 \pm 0.13 \deg$. 

We therefore find that the combined solar and anthropogenic sources were responsible for an increase of $0.50 \pm 0.35 \deg$. This should be compared with the observed $0.57 \pm 0.17 \deg$ increase in global surface temperature \citep{IPCC}. In other words, the result we find for the sensitivity and drivers are consistent with the observed temperature increase. The solar and anthropogenic contributions are comparable, nevertheless, it is more likely that the former contribution is somewhat larger, contributing about 2/3s of the observed temperature increase. This conclusion was independently reached by comparing the non-monotonic temperature increase with the non-monotonic solar activity increase and the monotonic increase in GHGs \citep{Soon96}.


We therefore conclude that both GHGs and solar variability are the main drivers of global warming, through a temperature sensitivity that is notably lower than the whole range obtained in GCMs. Its value is surprisingly close to the black body value of $0.30\CWm$.


\section*{\em Acknowledgments} 

The author wish the financial support of the ISF/Bikura fund for its support.







\end{article}
\newpage

\end{document}